\documentclass[letterpaper, 10 pt, conference]{ieeeconf}  % Comment this line out
\IEEEoverridecommandlockouts                              % This command is only
\usepackage{graphicx}
\usepackage{amsmath, amssymb}
\usepackage{mathtools}
\usepackage[T1]{fontenc}
\usepackage{lmodern}
\usepackage{newtxtext, newtxmath}
\usepackage{subcaption}
\usepackage[noend]{algpseudocode}
\usepackage{algorithm}
\usepackage{url}
\usepackage{cite}
\usepackage{multirow}
\usepackage{subcaption}
\usepackage{color}
\allowdisplaybreaks[4]
\mathtoolsset{showonlyrefs}

\newcommand{\pdev}[2]{{\frac{\partial {#1}}{\partial {#2}}}}
\newcommand{\pldev}[2]{{\partial {#1} / \partial {#2}}}
\newcommand{\bm}[1]{{\boldsymbol{\mathrm{#1}}}}
\newcommand{\D}{{\mathcal{D}}}
\newcommand{\Ore}{{\mathsf{O}}}
\newcommand{\shift}{{\mathsf{S}}}

\newcommand{\dop}[1]{{\mathcal{#1}}}
\newcommand{\sop}[1]{{\mathsf{#1}}}

\newtheorem{definition}{Definition}
\newtheorem{assumption}{Assumption}
\newtheorem{lemma}{Lemma}

\algrenewcommand\algorithmicrequire{\textbf{Input:}}
\algrenewcommand\algorithmicensure{\textbf{Output:}}
\algrenewcommand\algorithmicforall{\textbf{for each}}

\title{\LARGE \bf
Symbolic-Numeric Computation of Integrals in \\
Successive Galerkin Approximation of \\
Hamilton-Jacobi-Bellman Equation 
}

\author{Tomoyuki Iori% <-this % stops a space
\thanks{This work was partly supported by JSPS KAKENHI Grant Numbers JP21K21285 and 22K17855.}% <-this % stops a space
\thanks{T. Iori is with the Department of Information and Physical Sciences, Graduate School of Information Science and Technology, Osaka University, 1-5 Yamadaoka, Suita, Osaka 565--0871, Japan {\tt\small t-iori@ist.osaka-u.ac.jp}}%
}

\begin{document}

\maketitle
\thispagestyle{empty}
\pagestyle{empty}

%%%%%%%%%%%%%%%%%%%%%%%%%%%%%%%%%%%%%%%%%%%%%%%%%%%%%%%%%%%%%%%%%%%%%%%%%%%%%%%%
\begin{abstract}
	This paper proposes an efficient symbolic-numeric method to compute the integrals in the successive Galerkin approximation (SGA) of the Hamilton-Jacobi-Bellman (HJB) equation.
	A solution of the HJB equation is first approximated with a linear combination of the Hermite polynomials. 
	The coefficients of the combination are then computed by iteratively solving a linear equation, which consists of the integrals of the Hermite polynomials multiplied by nonlinear functions. 
	The recursive structure of the Hermite polynomials is inherited by the integrals, and their recurrence relations can be computed by using the symbolic computation of differential operators. 
	By using the recurrence relations, all the integrals can be computed from a part of them that are numerically evaluated. 
	A numerical example is provided to show the efficiency of the proposed method compared to a standard numerical integration method. 
\end{abstract}

%%%%%%%%%%%%%%%%%%%%%%%%%%%%%%%%%%%%%%%%%%%%%%%%%%%%%%%%%%%%%%%%%%%%%%%%%%%%%%%%
\section{Introduction \label{sec:intro}}

For general nonlinear systems, it is well-known that the optimal feedback controller with infinite horizon is provided by a solution of the Hamilton-Jacobi-Bellman (HJB) equation~\cite{Bryson1975}. 
In the case of linear systems with a quadratic performance index, the HJB equation reduces to a matrix equation called the Riccati equation, which can be solved efficiently. 
However, for nonlinear cases, the HJB equation is formulated as a nonlinear partial differential equation (PDE), which is extremely difficult to solve analytically. 
Hence, there have been many studies on numerical approximation techniques of a solution of the HJB equation. 

One possibility is to approximate a solution of the HJB equation by the Taylor-series approximation~\cite{Lukes1969,Garrard1972,Almubarak2019}.
By substituting the truncated Taylor-series expansion of a solution to the HJB equation, a finite number of algebraic equations for the coefficients of the truncated expansion can be obtained. 
Moreover, in~\cite{Almubarak2019}, a recursive closed form procedure for computing the coefficients is proposed under several conditions. 
% The stable manifold method is one of the most promising techniques, where  

The stable manifold method~\cite{Sakamoto2008} is one of the most promising techniques. 
In this method, a sequence of trajectories that converges to a trajectory on the stable manifold is iteratively computed. 
After sufficient number of iterations, the trajectory can be viewed as an approximation of the Hamiltonian flow on the stable manifold, and thus, it approximately gives the optimal feedback controller. 
% the associated Hamiltonian system is iteratively solved backward from a neighborhood of the origin so that its flow converges to a trajectory on the stable manifold as the iteration number increases. 

Another approach is to solve a linear PDE called the generalized HJB (GHJB) equation iteratively, which is called a successive approximation approach (SAA)~\cite{Beard1997}. 
As the GHJB equation is still difficult to solve, the GHJB equation and its solution is further approximated by using multiple basis functions, which is called the successive Galerkin approximation (SGA). 
It is proved in~\cite{Beard1997} that the solutions of the SGA converges to a solution of the HJB as the number of basis functions and iterations go to infinity. 
Several modified versions of the SAA have been studied~\cite{Mizuno2008,Kalise2018,Maruta2020}. 

One main issue of the SGA is the computation of integrals associated with the inner product of a function space. 
Even if all the integrals can be computed offline, their computational cost may still be too high to complete within a reasonable amount of computing power and time.  
Therefore, it is still important to seek an efficient method for computing the inner products or the integrals of nonlinear functions. 

In recent years, the symbolic computation of differential operators has been intensively studied~\cite{Saito2000,Oaku2003,Hibi2013,Oaku2018}, and its applications can be found in statistics~\cite{Nakayama2011,Kume2018}, moment problems~\cite{Brehard2019}. 
% It is also applied to the Bayesian filtering problem of nonlinear systems~\cite{Iori2020,Iori2021,Iori2022} and to the HJB equation from the perspective of first integrals~\cite{Iori2022}. 
In the theory of D-modules~\cite{Coutinho1995}, the solutions of PDEs are characterized by an infinite set of differential operators called an ideal.
It can be proven that any ideal is represented by a finite number of differential operators, and thus, it can be computed symbolically by using the computer algebra systems (CASs) such as~\texttt{Singular}~\cite{Singular}, \texttt{Macaulay2}~\cite{Macaulay2}, and \texttt{Risa/Asir}~\cite{NoroAsir}.
% It can be represented by a finite number of differential operators and can be computed symbolically by using the computer algebra systems (CASs) such as~\cite{Singular}, \cite{Macaulay2}, and \cite{NoroAsir}. 
In particular, a specific type of ideals called holonomic ideals and their solutions called holonomic functions are the main objects manipulated by the symbolic computation. 
More specifically, for the integral of a holonomic function, a holonomic ideal that the integral satisfies can be computed exactly from a holonomic ideal that the integrand satisfies. 
This provides a great advantage in the design of algorithms because the integral of a general nonlinear function cannot be computed exactly. 
In addition to differential equations, recurrence equations can also be treated by the symbolic computation by converting them to differential equations via the Mellin transform~\cite{Oaku2003,Oaku2018}.

In this paper, we propose a method to compute recurrence relations satisfied by the integrals required for the SGA.  
% We first assume that the plant dynamics is described by holonomic functions. 
The expansion of a solution of the HJB equation with Hermite polynomials, which is holonomic as an infinite sequence of functions is considered. 
In addition, the plant dynamics is assumed to be described by holonomic functions. 
These problem settings allow us to compute a set of recurrence relations that are satisfied by the integrals for the SGA. 
Once the recurrence equations are computed, any number of the integrals can be readily computed just by substituting several initial values to the recurrence equations. 
Moreover, in contrast to the earlier methods~\cite{Mizuno2008,Kalise2018}, the proposed method can consider the nonlinearities of the plant dynamics exactly in the computation of the integrals. 

\paragraph*{Notations}
Let $\bm{N}$ be the set of positive integers and $\bm{R}$ be the field of real numbers. 
For $(i_1,\dots,i_p) \in \bm{N}^p$ and $(x_1,\dots,x_n) \in \bm{R}^n$, $F(i; x) = F(i_1,\dots,i_p; x_1,\dots,x_n)$ denotes a function that is defined on a subset of $\bm{N}^p \times \bm{R}^n$. 
If $p = 0$ or $n = 0$, $F(i; x)$ or $F(i; x)$ are simply written as $F(i)$ or $F(x)$, respectively. 
The symbols $\shift_{i_j}, \shift_{i_j}^{-1}\ (j \in \{1,\dots,p\})$, and $\partial_{x_j}\ (j \in \{1,\dots,p\})$ denote the shift operators for $i_j$ and the differential operator for $x_j$, respectively. 
That is, $\shift_{i_j} \bullet F(i; x) = F(i_1,\dots,i_j+1,\dots,i_p; x)$, $\shift_{i_j} \bullet F(i; x) = F(i_1,\dots,i_j-1,\dots,i_p; x)$, and $\partial_{x_j} \bullet F(i; x) = \pldev{F}{x_j}(i; x)$, where $\bullet$ denotes the action of a differential or difference operator on a function. 
% For a vector of indeterminates $y = [y_1\ \cdots\ y_n]^\T$, $\bm{R}(y)$ denotes the field of rational functions in $y_1,\dots,y_n$ over $\bm{R}$. 
% For a vector of indeterminates $y = [y_1\ \cdots\ y_n]^\T$, $\bm{R}[y] = \bm{R}[y_1,\dots,y_n]$ denotes the commutative ring of polynomials in $y$. 
% $\partial_{y_i}$ and $\shift_{y_i}$ denote the differential operator and shift operator, respectively. 
% $\partial_y \coloneqq [\partial_{y_1}\ \cdots\ \partial_{y_n}]^\T$ denotes a vector of differential operators, where $\partial_{y_i} = \pldev{}{y_i}$. 
% Here, $\partial_y$ and $\partial_{y_i}$ are abbreviated by $\partial$ and $\partial_i$, respectively, if $y$ is clearly specified according to the context. 
Let $\bm{R}[x]$ be the set of all polynomials in $x_1,\dots,x_n$ with coefficients in $\bm{R}$. 
The symbol $\bm{R}[x]\langle \partial_x \rangle$ denotes the noncommutative ring of differential operators with coefficients in $\bm{R}[x]$. 
Similarly, $\bm{R}[i]\langle \shift_i \rangle$ denotes the noncommutative ring of difference operators with coefficients in $\bm{R}[i]$. 
The symbols $\bm{R}[x]\langle \partial_x \rangle$ and $\bm{R}[i]\langle \shift_i \rangle$ are also denoted by $\D_n$ and $\Ore_p$, respectively, if the indeterminates $x$ and $i$ are clearly specified according to the context. 
If $\dop{P} \bullet f = 0$ for $\dop{P} \in \D_n$, $\dop{P}$ is said to \emph{annihilate} a function $f$ and $f$ is a \emph{solution} of $\dop{P}$. 
If $\sop{E} \bullet f = 0$ for $\sop{E} \in \Ore_p$, $\sop{E}$ is said to \emph{annihilate} $f$ and $f$ is a \emph{solution} of $\sop{E}$. 

\section{Problem Setting}
% Consider the following control-affine nonlinear system with a scalar state: 
Consider the optimal control problem of a nonlinear system:
\begin{equation}
	\dot{x} = f(x) + g(x)u \label{eq:sys}
\end{equation}
with a performance index defined as
\begin{equation}
	% J(x, u) \coloneqq \int_0^{\infty} q(x(t)) + ru(t)^2 dt, 
	J(x, u) \coloneqq \int_0^{\infty} q(x(t)) + \|u(t)\|^2_R dt, 
\end{equation}
where $x \in \bm{R}^n$ and $u \in \bm{R}^m$ are the state and input of the system, respectively. 
The nonlinear functions $f\colon \bm{R}^n \to \bm{R}^n$, $g\colon \bm{R}^n \to \bm{R}^{n \times m}$, and $q\colon \bm{R}^n \to \bm{R}$ are assumed to consists of holonomic functions, which will be defined in Section~\ref{sec:holonomic}. 
It is also assumed that $q(x) \geq 0\ (x \in \bm{R}^n)$ and $R \in \bm{R}^{m \times m}$ be positive definite. 

% The nonlinear functions $f, g\colon \bm{R} \to \bm{R}$ with $f(0) = 0$ are assumed to be holonomic~\cite{Kauers2013,Hibi2013}. 
% The performance index for this controlled system is defined as
% \begin{equation}
% 	J \coloneqq \int_0^{\infty} q(x(t)) + ru(t)^2 dt, 
% \end{equation}
% where $q\colon \bm{R} \to \bm{R}$ is positive definite and $r \in \bm{R}$ is a positive scalar weight. 

% Suppose the input
% Suppose the input is given by a feedback control law $u(x)$. 
% By letting $\phi(t)$ be the solution of~\eqref{eq:sys} with an initial state $\phi(0) = x_0 \in \bm{R}$ and $u(x)$, the performance index is defined as
% For system~\eqref{eq:sys}, consider the problem of finding a feedback control law $u(x)$ that minimizes the performance index
% % A performance index for system~\eqref{eq:sys} is defined as
% \begin{equation}
% 	V(x_0, u(\cdot)) \coloneqq \int_0^{\infty} q(\phi(t)) + ru(\phi(t))^2 dt, 
% \end{equation}
% where $q\colon \bm{R} \to \bm{R}$ is positive definite, $r \in \bm{R}$ is a positive scalar weight, and $\phi(t)$ denotes the solution of~\eqref{eq:sys} with the initial state $\phi(0) = x_0$ and the feedback law $u(\cdot)$. 

% The theory of the dynamic programming shows that the optimal feedback law $u^*(x)$ and optimal value function $V^*(x) \coloneqq V(x, u^*(x))$ satisfy the following Hamilton-Jacobi-Bellman (HJB) equation:  
% \begin{equation}
% 	0 = q(x) + \min_{u(\cdot)} \left\{ \pdev{V^*}{x}(f + Gu) + \| u(x) \|_R \right\}. 
% \end{equation}
For a given feedback law $u(x)$, the performance index for the closed loop trajectory $\phi(t)$ is defined as
\begin{equation}
	V(x_0, u(\cdot)) \coloneqq \int_0^{\infty} q(\phi(t)) + \|u(\phi(t))\|^2_R dt, 
\end{equation}
where $\phi(0) = x_0$. 
The optimal feedback law $u^*(x)$ is given as
\begin{equation}
	% u^*(x) = -\frac{1}{2r}g(x)\pdev{V^*}{x}(x), \label{eq:V2u}
	u^* = -\frac{1}{2}R^{-1}g^\top\pdev{V^*}{x}^\top, \label{eq:V2u}
\end{equation}
where $V^*\colon \bm{R}^n \to \bm{R}$ is a solution of the HJB equation
\begin{equation}
	% \mathrm{HJB}(V^*) \coloneqq \pdev{V^*}{x}f + q -\frac{1}{4r}g^2\pdev{V^*}{x}^2 = 0, 
	\mathrm{HJB}(V^*) \coloneqq \pdev{V^*}{x}f + q -\frac{1}{4}\pdev{V^*}{x}gR^{-1}g^\top\pdev{V^*}{x}^\top = 0, 
\end{equation}
such that $V^*(0) = 0$.
%  and the autonomous system $\dot{x} = f + gu^*$ is asymptotically stable. 

% \section{Successive Galerkin Approximation}

% Successive Galerkin approximation (SGA) is an algorithm to approximate the value function successively. 

% Successive approximation approach~\cite{}
In the SAA, a solution of the HJB equation is iteratively approximated by solving the GHJB equation defined as
% For a given feedback law $u(x)$, the generalized HJB equation is defined as
\begin{equation}
	\mathrm{GHJB}(V, u) \coloneqq \pdev{V}{x}(f + gu) + q + \|u\|^2_R = 0. 
\end{equation}
% More specifically, for a given initial guess $u^{(0)}$, $V^*$ can be approximated iteratively by the following algorithm.
% \begin{enumerate}
% 	\item $l \gets 0$
% 	\item Solve $\mathrm{GHJB}(V^{(l)}, u^{(l)}) = 0$ to compute $V^{(l)}(x)$ \label{enum:u2V}
% 	\item Define $u^{(l+1)}$ by~\eqref{eq:V2u} with $V^{(l)}$
% 	\item $l \gets l+1$ and go back to Step \ref{enum:u2V}
% \end{enumerate}
As the GHJB equation is still difficult to solve, the iteration $V^{(l)}$ is further approximated by using a finite number of basis functions $\phi_1(x),\dots,\phi_N(x)$ as
% As the GHJB equation is still difficult to solve, the iteration $V^{(l)}$ is further approximated by using a finite number of basis functions.
% In this paper, $V^{(l)}$ is approximated with a linear combination of the Hermite polynomials $H(1; x), \dots, H(N; x)$ up to degree $N$ as follows. 
\begin{equation}
	V^{(l)}(x) \coloneqq \sum_{i=1}^N v_i^{(l)} \phi_i(x) = \Phi^\top(x) v^{(l)} \quad \left(v_i^{(l)} \in \bm{R}\right), \label{eq:Vapprox}
	% V^{(l)}(x) \coloneqq \sum_{i=1}^N v_i^{(l)} H(i; x) = \Phi^\top(x) v^{(l)} \quad \left(v_i^{(l)} \in \bm{R}\right), \label{eq:Vapprox}
\end{equation}
where $\Phi(x) \coloneqq [\phi_1(x)\ \cdots\ \phi_N(x)]^\top$ and $v^{(l)} \coloneqq [v_1^{(l)}\ \cdots\ \allowbreak v_N^{(l)}]^\top$. 
% where $\Phi(x) \coloneqq [H(1; x)\ \cdots\ H(N; x)]^\top$ and $v^{(l)} \coloneqq [v_1^{(l)}\ \cdots\ v_N^{(l)}]^\top$. 
This also yields a representation of $u^{(l+1)}$ as
\begin{align}
	% u^{(l)}(x) = -\frac{1}{2r}g(x)\pdev{V^{(l)}}{x}(x) = \sum_{i = 1}^N v_i^{(l-1)}\left\{ -\frac{1}{2r}g(x)\pdev{\phi_i}{x} \right\}, 
	% u^{(l)}(x) = -\frac{1}{2r}g(x)\pdev{V^{(l)}}{x}(x) = -\frac{1}{2r}g(x)\pdev{\Phi^\top}{x}(x) v^{(l)}, 
	u^{(l+1)}(x) &= -\frac{1}{2}R^{-1}g^\top(x)\pdev{V^{(l)}}{x}^\top(x) \\
	&= -\frac{1}{2}R^{-1}g^\top(x)\pdev{\Phi}{x}^\top(x) v^{(l)}. \label{eq:uapprox}
\end{align}
The functional equation $\mathrm{GHJB}(V^{(l)}, u^{(l)}) = 0$ is then approximated with the following finite number of algebraic equations for $v_i^{(l)}$: 
\begin{equation}
	\left\langle \mathrm{GHJB}(V^{(l)}, u^{(l)}), \phi_k \right\rangle = 0 \quad (p = 1,\dots,N), \label{eq:GHJBIP}
\end{equation}
where $\langle \cdot, \cdot \rangle$ is an inner product of functions defined as
\begin{equation}
	\langle f(x), g(x) \rangle \coloneqq \int_{\bm{R}^n} w(x) f(x) g(x) dx \label{eq:IP}
\end{equation}
with a weighting function $w(x) > 0\ (x \in \bm{R}^n)$. 
By substituting~\eqref{eq:Vapprox} and~\eqref{eq:uapprox} into~\eqref{eq:GHJBIP}, we obtain
% \eqref{eq:GHJBIP} can be rewritten as follows:
\begin{align}
	% &\left\langle \mathrm{GHJB}(V^{(l)}, u^{(l)}), \phi_k \right\rangle \\
	% =& \left\langle \sum_{i = 1}^N v_i^{(k)} \pdev{\phi_i}{x}f, \phi_p \right\rangle + \left\langle	\sum_{i=1}^N v_i^{(k)} \pdev{\phi_i}{x}G \sum_{j=1}^N v_j^{(k-1)}\left\{ -\frac{1}{2}R^{-1}G^\top\pdev{\phi_j}{x}^\top \right\}, \phi_p \right\rangle \\
	% &+ \left\langle q, \phi_p \right\rangle + \left\langle \left\| \sum_{j=1}^N v_j^{(k-1)}\left\{ -\frac{1}{2}R^{-1}G^\top\pdev{\phi_j}{x}^\top \right\} \right\|_{R}^2, \phi_p \right\rangle \\
	% =& \sum_{i = 1}^N v_i^{(k)} \left\langle \pdev{\phi_i}{x}f, \phi_p \right\rangle \\
	&\left\langle \mathrm{GHJB}(V^{(l)}, u^{(l)}), \phi_k \right\rangle = \sum_{i = 1}^N v_i^{(l)} \left\langle \pdev{\phi_i}{x}f, \phi_k \right\rangle \\
	&- \frac{1}{2} \sum_{i=1}^N v_i^{(l)} \sum_{j=1}^N v_j^{(l-1)} \left\langle \pdev{\phi_i}{x}G R^{-1}G^\top\pdev{\phi_j}{x}^\top, \phi_k \right\rangle \\
	&+ \left\langle q, \phi_k \right\rangle +\frac{1}{4} \sum_{i=1}^N v_i^{(l-1)} \sum_{j=1}^N v_j^{(l-1)}\left\langle\pdev{\phi_i}{x}G R^{-1}G^\top\pdev{\phi_j}{x}^\top, \phi_k \right\rangle. 
\end{align}
% \begin{align}
% 	&\left\langle \mathrm{GHJB}(V^{(l)}(x), u^{(l)}(x)), \phi_k(x) \right\rangle \\
% 	=& \left\langle \sum_{i = 1}^N v_i^{(l)} \pdev{\phi_i}{x}f, \phi_k \right\rangle + \left\langle	\sum_{i=1}^N v_i^{(l)} \pdev{\phi_i}{x}g \sum_{j=1}^N v_j^{(l-1)}\left\{ -\frac{1}{2r}g\pdev{\phi_j}{x} \right\}, \phi_k \right\rangle \\
% 	&+ \left\langle q, \phi_k \right\rangle + \left\langle r\left\{ \sum_{j=1}^N v_j^{(l-1)}\left( -\frac{1}{2r}g\pdev{\phi_j}{x}^\top \right) \right\}^2, \phi_k \right\rangle \\
% 	=& \sum_{i = 1}^N v_i^{(l)} \left\langle \pdev{\phi_i}{x}f, \phi_k \right\rangle - \frac{1}{2} \sum_{i=1}^N \sum_{j=1}^N v_i^{(l)} v_j^{(l-1)} \left\langle \frac{1}{r}g^2\pdev{\phi_i}{x}\pdev{\phi_j}{x}, \phi_k \right\rangle \\
% 	&+ \left\langle q, \phi_k \right\rangle +\frac{1}{4} \sum_{i=1}^N \sum_{j=1}^N v_i^{(l-1)} v_j^{(l-1)}\left\langle \frac{1}{r}g^2\pdev{\phi_i}{x}\pdev{\phi_j}{x}, \phi_k \right\rangle. 
% \end{align}
Now, let us define
\begin{equation}
	% v^{(l)} &\coloneqq \left[v_1^{(l)}\ \cdots\ v_N^{(l)}\right]^\top, \\
	\begin{gathered}
	a(i, j, k) \coloneqq \left\langle\pdev{\phi_i}{x}G R^{-1}G^\top\pdev{\phi_j}{x}^\top, \phi_k \right\rangle, \\
	b(i, k) \coloneqq \left\langle \pdev{\phi_i}{x}f, \phi_k \right\rangle, \quad c(k) \coloneqq \left\langle q(x), \phi_k(x) \right\rangle. 
	\end{gathered} \label{eq:integrals}
\end{equation}
% \begin{align}
% 	% v^{(l)} &\coloneqq \left[v_1^{(l)}\ \cdots\ v_N^{(l)}\right]^\top, \\
% 	a[i, j, k] &\coloneqq \left\langle \frac{1}{r}g^2(x)\pdev{\phi_i}{x}(x)\pdev{\phi_j}{x}(x), \phi_k(x) \right\rangle, \label{eq:IP1}\\
% 	b[i, k] &\coloneqq \left\langle \pdev{\phi_i}{x}(x)f(x), \phi_k(x) \right\rangle, \label{eq:IP2}\\
% 	c[k] &\coloneqq \left\langle q(x), \phi_k(x) \right\rangle, \label{eq:IP3}
% \end{align}
In addition, let $A_k$ be the $N \times N$ matrix having $a(i, j, k)$ as its $(i,j)$-component and $b_k$ be the $N$-dimensional vector having $b(i, k)$ as its $i$-th component. 
The approximated GHJB equation~\eqref{eq:GHJBIP} is then rewritten as the following linear equations for $v^{(l)}$: 
\begin{multline}
	\mathrm{LE}(v^{(l)}, v^{(l-1)}) \coloneqq \\
	{v^{(l)}}^\top \left( b_k -\frac{1}{2} A_kv^{(l-1)}  \right) + c(k) + \frac{1}{4} \left\| v^{(l-1)} \right\|_{A_k}^2 = 0 \\
	(k=1,\dots,N). \label{eq:LE}
\end{multline}
% In what follows, the parameters $a[i,j,k]$, $b[i,k]$, and $c[k]$ are called the \textit{SGA parameters}. 
In what follows, for a given $N \in \bm{N}$, $A_N$, $B_N$, and $C_N$ denote the sets of the integrals required to define~\eqref{eq:LE}, that is, 
\begin{align}
	&A_N \coloneqq \{a(i,j,k) \mid i,j,k \in \{1,\dots,N\}\}, \\
	&B_N \coloneqq \{b(i,k) \mid i,k \in \{1,\dots,N\}\}, \\
	&C_N \coloneqq \{c(k) \mid k \in \{1,\dots,N\}\}. 
\end{align}

Once the integrals $A_N$, $B_N$, and $C_N$ are computed, the SGA procedure can be performed by solving the linear equation~\eqref{eq:LE} iteratively, which is summarized as Algorithm~\ref{alg:SGA}. 
\begin{algorithm}[b]
	\caption{Successive Galerkin approximation\label{alg:SGA}}
	\begin{algorithmic}[1]
		\Require{Initial guess $v^{(0)}$, sets of integrals $A_N$, $B_N$, and $C_N$, error tolerance $\epsilon$, max iteration $l_{\mathrm{max}}$}
		\Ensure{Approximation of optimal value function $V^{(l)}(x) = \Phi^\top(x)v^{(l)}$}
		\State{$l \gets 0$}
		\While{$\mathrm{LE}(v^{(l)}, v^{(l)}) > \epsilon$ and $l < l_{\mathrm{max}}$}
		\State{Solve linear equations~\eqref{eq:LE} and compute $v^{(l+1)}$}
		\State{$l \gets l+1$}
		\EndWhile
		\State\Return{$V^{(l)} = \Phi^\top(x)v^{(l)}$}
		% \If{$\mathrm{LE}(v^{(l)}, v^{(l)})$ is sufficiently small or $l > l_{\mathrm{max}}$}
		% \State{$v^{(\infty)} \gets v^{(l+1)}$}
		% \State\Return{$v^{(\infty)}$}
		% \EndIf
		% \State{}
	\end{algorithmic}
\end{algorithm}

% The SGA parameters never be computed again in the iteration. 
% However, the coefficients are defined as integrals, which are extremely difficult to compute analytically.
Although $A_N$, $B_N$, and $C_N$ are never computed again in the iteration, they are defined as integrals of nonlinear functions, which are extremely difficult to compute analytically. 
Hence, we have to rely on numerical integration methods, which leads to high computational cost. 
% Numerical integration is one remedy to this issue, but it will lead to high computational cost, which is unavailable even if it can be computed offline. 
% Consequently, the development of more efficient algorithms to compute $A_N$, $B_N$, and $C_N$ is important. 
However, if the basis functions have some recurrence relations such as those of the orthogonal polynomials, we can expect that the elements of $A_N$, $B_N$, and $C_N$ may also have similar recursive relations. 
Such recursive relations can be computed by using the symbolic computation of differential operators, which will be briefly introduced in the next section. 

%%%%%%%%%%%%%%%%%%%%%%%%%%%%%%%%%%%%%%%%%%%%%%%%%%%%%%%%%%%%%%%%%%%%%%%%%%

\section{Holonomic Functions and Ideals\label{sec:holonomic}}

This section is devoted to introducing some definitions and lemmas related to the symbolic computation of differential operators, referring to~\cite{Oaku2003,Hibi2013}. 
In this section, $x = [x_1\ \cdots\ x_n]^\top \in \bm{R}^n$, $s = [s_1\ \cdots\ s_p]^\top \in \bm{R}^p$, and $i = [i_1\ \cdots\ i_p]^\top \in \bm{Z}^p$.
Let $\D_n$ and $\D_{n+p}$ denote $\bm{R}[x]\langle \partial_x \rangle$ and $\bm{R}[x,s]\langle \partial_x, \partial_s \rangle$, respectively. 
Furthermore, let $\D_n\langle i, \shift_i, \shift_i^{-1} \rangle$ denote $\D_n$-algebra generated by $i$, $\shift_i = [\shift_{i_1}\ \cdots\ \shift_{i_p}]^\top$, and $\shift_i^{-1} = [\shift_{i_1}^{-1}\ \cdots\ \shift_{i_p}^{-1}]^\top$. 

% Holonomic functions are defined through the differential operators that annihilate . 
Consider a set of PDEs for an unknown function $f(x)$: 
\begin{equation}
	\dop{P}_1 \bullet f = \cdots \dop{P}_d \bullet f = 0, \label{eq:PDE}
\end{equation}
where $\dop{P}_1,\dots,\dop{P}_d \in \D_n$. 
For any solution $f$ of~\eqref{eq:PDE} and any differential operator $\dop{Q} \in \D_n$, it is obvious that 
$\dop{Q}\dop{P}_j \bullet f = \dop{Q} \bullet 0 = 0$, 
% \begin{equation}
% 	\dop{Q}\dop{P}_j \bullet f = \dop{Q} \bullet 0 = 0, 
% \end{equation}
which leads to consider a left ideal $I \subset \D_n$ defined as 
$I \coloneqq \{\dop{Q}_1\dop{P}_1 + \cdots \dop{Q}_d\dop{P}_d \in \D_n \mid \dop{Q}_1,\dots,\dop{Q}_d \in \D_n\}$. 
% $I \coloneqq \{\dop{P}\}$
% \begin{multline}
% 	I \coloneqq \{\dop{Q}_1\dop{P}_1 + \cdots \dop{Q}_d\dop{P}_d \in \D_n \mid \dop{Q}_1,\dots,\dop{Q}_d \in \D_n\}. 
% \end{multline}
In this paper, the adjective ``left'' is omitted because all ideals in this paper are left ideals. 
\begin{definition}[Holonomic ideal of $\D_n$\label{def:holideal}]
	An ideal $I \subset \D_n$ is holonomic if the quotient $\D_n/I$, which can be viewed as a left $\D_n$-module, has the dimension $n$. 
\end{definition}

For the definition of the dimension of left $\D_n$-modules, see~\cite{Coutinho1995,Saito2000,Hibi2013}. 
The notion of holonomic ideals plays an important role in the symbolic computation of differential operators because of its finiteness~\cite{Saito2000}. 

The solutions of holonomic ideals are called holonomic functions and include most nonlinear functions appears in systems and control theory~\cite{Koutschan2009,Iori2022}. 
\begin{definition}[Holonomic function\label{def:holfunc}]
	% An analytic function $f(x)$ is said to be holonomic if there exists a holonomic ideal $I \subset \D_n$ that annihilates $f$, that is, $\dop{P}\bullet f = 0$ for all $\dop{P} \in I$.
	An analytic function $f(x)$ is said to be holonomic if $f$ is annihilated by a holonomic ideal $I \subset \D_n$, that is, $\dop{P}\bullet f = 0$ for all $\dop{P} \in I$.
\end{definition}

The class of holonomic functions is closed under multiplication and integration; the product of two holonomic functions and the integral of a holonomic function are also holonomic\cite{Oaku2003}. 
\begin{lemma}\label{lem:holprod}
	For two holonomic functions $f(x)$ and $g(x)$, their product $(f \cdot g)(x)$ is holonomic. 
\end{lemma}
\begin{lemma}\label{lem:holinteg}
	Suppose a holonomic function $f(x)$ is \textit{rapidly decreasing} with respect to $x_1$, that is, $\lim_{x_1 \to \infty} x_1^i\partial_{x_1}^jf(x) = 0$ for any nonnegative integers $i,j$. 
	Then, the integral $\int_{-\infty}^{\infty} f(x) dx_1$ is holonomic as a function of $[x_2,\dots,x_n]^\top$. 
\end{lemma}

% From an infinite sequence of functions $f[i](x)$, we can consider a nonlinear function $\tilde{f}(s, x)$ that is related to the infinite sequence by the Mellin transform. 
% \begin{definition}
% 	The \textit{inverse Mellin transform} $\tilde{f}(s, x)$ of $f[i](x)$ is defined as
% 	\begin{equation}
% 		\tilde{f}(s, x) \coloneqq \sum_{i_1=0}^{\infty} \cdots \sum_{i_p=0}^{\infty} f[i](x) s_1^{-i_1} \cdots s_p^{-i_p}, 
% 	\end{equation}
% 	if the summation on the right-hand side converges. 
% \end{definition}
Recurrence relations of infinite sequences can be converted into PDEs and manipulated by the symbolic computation of differential operators through the Mellin transform. 
\begin{definition}[Mellin transform between $\partial$ and $\shift$] \label{def:mellin}
	% The \textit{Mellin transform} $\mu\colon \D_{n+p} \to \D_n\langle i, \shift_i, \shift_i^{-1} \rangle$ is defined as the following $\D_n$-algebra homomorphism: 
	A mapping $\mu\colon \D_{n+p} \to \D_n\langle i, \shift_i, \shift_i^{-1} \rangle$ defined as
	\begin{equation}
		\mu(s_j) = \shift_{i_j}, \quad \mu(\partial_{s_j}) = -i_j\shift_{i_j}^{-1} \label{eq:MT}
	\end{equation}
	is called the \textit{Mellin transform}, 
	and a mapping $\hat{\mu}\colon \D_n\langle i, \shift_i \rangle\allowbreak \to \D_{n+p}$ defined as
	\begin{equation}
		\hat{\mu}(i_j) = -\partial_{s_j}s_j = -s_j\partial{s_j} - 1, \quad \hat{\mu}(\shift_{i_j}) = s_j \label{eq:IMT}
	\end{equation}
	is called the \textit{inverse Mellin transform}. 
\end{definition}

% The Mellin transform is related to the transformation $\mathscr{M}$ between a multivariate function and an infinite sequence of functions: 
% \begin{align}
% 	\mathscr{M}[g](i; x) &\coloneqq \int_{C} g(s, x)s_1^{i_1} \cdots s_p^{i_p}ds, \\
% 	\mathscr{M}^{-1}[f](s, x) &\coloneqq \sum_{i_1=0}^{\infty} \cdots \sum_{i_p=0}^{\infty} f(i; x)s_1^{-i_1-1} \cdots s_p^{-i_p-1}, 
% \end{align}
The Mellin transform is related to the multivariate z-transformation $\mathscr{Z}$ between a multivariate function and an infinite sequence of functions: 
\begin{align}
	\mathscr{Z}[f](s, x) &\coloneqq \sum_{i_1=0}^{\infty} \cdots \sum_{i_p=0}^{\infty} f(i; x)s_1^{-i_1-1} \cdots s_p^{-i_p-1}, \\
	\mathscr{Z}^{-1}[g](i; x) &\coloneqq \int_{C} g(s, x)s_1^{i_1} \cdots s_p^{i_p}ds, 
\end{align}
where $C$ is the product of unit circles centered at the origin of the complex plane. 
It is readily seen that
\begin{equation}
	\hat{\mu}(i) \bullet \mathscr{Z}[f] = \mathscr{Z}[if], \quad \hat{\mu}(\shift_i) \bullet \mathscr{Z}[f] = \mathscr{Z}[\shift_if].
\end{equation}

\section{Recurrence Equations of Integrals \label{sec:RecRel}}

% Hereafter, suppose $n = 1$the $i$-th basis function $\phi_i(x)$ in~\eqref{eq:Vapprox} is the Hermite polynomial $H[i](x)$ of degree $i$.
Hereafter, suppose $n = 1$ for simplicity and that the $i$-th basis function $\phi_i(x)$ in~\eqref{eq:Vapprox} is the Hermite polynomial $H(i; x)$ of degree $i$. 
For multi-dimensional cases, each basis function can be defined as the product of the Hermite polynomials $H(i_1; x_1)H(i_2; x_2) \cdots H(i_n; x_n)$, and the following discussions can be applied to with appropriate modifications, which will be a part of future work. 

% The integrals~\eqref{eq:integrals} are rewritten as follows. 
% \begin{align}
% 	a(i,j,k) &= \left\langle \frac{1}{R}g^2(x) \partial_xH(i; x) \partial_xH(j; x), H(k; x) \right\rangle, \label{eq:HIP1}\\
% 	b(i,k) &= \left\langle (\partial_x H(i; x))f(x), H(k; x) \right\rangle, \label{eq:HIP2}\\
% 	c(k) &= \left\langle q(x), H(k; x) \right\rangle, \label{eq:HIP3}
% \end{align}
% where the weighting function is set to $w(x) = \exp(-x^2)$, and their z-transforms are obtained as follows. 
% Now, let us consider the z-transform of~\eqref{eq:HIP1}--\eqref{eq:HIP3} as follows. 
With the weighting function $w(x) = \exp(-x^2)$, the z-transforms of integrals~\eqref{eq:integrals} are obtained as follows.
\begin{align}
	% &\mathscr{Z}[a](s, t, v) = \sum_{i=0}^{\infty}\sum_{j=0}^{\infty}\sum_{k=0}^{\infty} a[i,j,k]s^{-i-1}t^{-j-1}u^{-k-1} \\
	% &= \left\langle \frac{1}{r}g^2(x)\mathscr{Z}[\partial_x H](s, x) \mathscr{Z}[\partial_x H](t, x), \mathscr{Z}[H](u, x) \right\rangle \\
	% &= \frac{1}{r} \int_{-\infty}^{\infty} \Big\{e^{-x^2}g^2(x)\mathscr{Z}[\partial_x H](s, x) \\
	\mathscr{Z}[a](s, t, v) &= \int_{-\infty}^{\infty} \frac{1}{R} \Big\{e^{-x^2}g^2(x)\mathscr{Z}[\partial_x H](s, x) \\
	&\qquad \qquad \mathscr{Z}[\partial_x H](t, x) \mathscr{Z}[H](u, x)\Big\}dx, \label{eq:ZT1}\\
	\mathscr{Z}[b](s, v) &= \int_{-\infty}^{\infty} e^{-x^2}\mathscr{Z}[\partial_x H](s, x) f(x) \mathscr{Z}[H](u, x)dx, \label{eq:ZT2}\\
	\mathscr{Z}[c](v) &= \int_{-\infty}^{\infty} e^{-x^2}q(x) \mathscr{Z}[H](u, x)dx. \label{eq:ZT3}
	% &\begin{multlined}
	% 	= \frac{1}{r} \int_{-\infty}^{\infty} \Big\{e^{-x^2}g^2(x)\mathscr{Z}[\partial_x H](s, x) \\
	% 	\mathscr{Z}[\partial_x H](t, x) \mathscr{Z}[H](u, x)\Big\}dx
	% \end{multlined}
\end{align}
If all the factors of the integrands in~\eqref{eq:ZT1}--\eqref{eq:ZT3} are holonomic, these z-transforms are also holonomic. 
Moreover, for each of~\eqref{eq:ZT1}--\eqref{eq:ZT3}, some differential operators that annihilates it can be computed by using the symbolic computation (for detailed algorithms, see~\cite{Saito2000,Oaku2003,Hibi2013}). 
Accordingly, we can obtain some recurrence relations form the differential operators via the Mellin transform~\eqref{eq:MT}. 
% , that is, 
% \begin{equation}
% 	\phi_i(x) = H_i(x) \coloneqq (-1)^n \exp(x^2) \partial_x^i \exp(-x^2). 
% \end{equation}

To this end, we next show that the z-transforms of the Hermite polynomials and their derivatives are holonomic. 
The Hermite polynomials satisfy the following difference-differential equations: 
\begin{align}
	(\partial_x^2 -2x\partial_x + 2i) \bullet H(i; x) = 0, \label{eq:ddohp1}\\
	(\shift_i^2 - 2x\shift_i + 2(i+1)) \bullet H(i; x) = 0. \label{eq:ddohp2}
\end{align}
By the inverse Mellin transform~\eqref{eq:IMT}, the following two differential operators are obtained. 
\begin{align}
	(\partial_x^2 - 2x\partial_x - 2s\partial_s - 2) \bullet \mathscr{Z}[H](s, x) = 0, \label{eq:dohp1}\\
	(s^2 - 2xs - 2s\partial_s) \bullet \mathscr{Z}[H](s, x) = 0. \label{eq:dohp2}
\end{align}
% where $H(s, x)$ is the inverse Mellin transform of $H(i; x)$.
% Moreover, it can be readily verified that \eqref{eq:dohp1} and \eqref{eq:dohp2} generate a holonomic ideal by using algorithms implemented in some modules of CASs such as \texttt{dmod.lib} for \texttt{Singular}~\cite{Singular} and \texttt{nk\_restriction.rr} for \texttt{Risa/Asir}~\cite{NoroAsir}. 
% Moreover, it can be readily verified that \eqref{eq:dohp1} and \eqref{eq:dohp2} generate a holonomic ideal by using the CASs such as \texttt{dmod.lib} for \texttt{Singular}~\cite{Singular} and \texttt{nk\_restriction.rr} for \texttt{Risa/Asir}~\cite{NoroAsir}. 
Moreover, it can be readily verified that \eqref{eq:dohp1} and \eqref{eq:dohp2} generate a holonomic ideal by using a CAS.
%  such as \texttt{Singular}, \texttt{Macaulay2}, and \texttt{Risa/Asir}.  
Hence, from Definition~\ref{def:holfunc}, $\mathscr{Z}[H](s, x)$ is holonomic. 
% Hence, from Definition~\ref{def:holideal_shift}, \eqref{eq:ddohp1} and \eqref{eq:ddohp2} generates a holonomic ideal of $\D \langle i, \shift_i \rangle$, and thus $H(i; x)$ is holonomic as a function of $i \in \bm{N}$ and $x \in \bm{R}$. 
\begin{lemma}
	The z-transform of the Hermite polynomials $\mathscr{Z}[H](s, x)$ is holonomic. 
\end{lemma}
% \begin{lemma}
% 	The Hermite polynomials $H(i; x)$ is holonomic in the sense of Definition~\ref{def:holfuncseq}. 
% \end{lemma}

As the derivative $\partial_x H(i; x)$ also appears in~\eqref{eq:ZT1} and~\eqref{eq:ZT2}, we next show that $\partial_x H(i; x)$ is holonomic.
To this end, the following equality satisfied by $H(i; x)$ is useful:
\begin{equation}
	\partial_x H(i; x) = 2iH(i-1; x). 
\end{equation}
From~\eqref{eq:ddohp1}, 
$(\partial_x^2 -2x\partial_x + 2(i-1)) \bullet H(i-1; x) = 0$
% \begin{equation}
% 	(\partial_x^2 -2x\partial_x + 2(i-1)) \bullet H(i-1; x) = 0 
% \end{equation}
holds. 
As $2i$ and $\partial_x^2 -2x\partial_x + 2(i-1)$ commute, we have
\begin{align}
	0 &= 2i(\partial_x^2 -2x\partial_x + 2(i-1)) \bullet H(i-1; x) \\
	&= (\partial_x^2 -2x\partial_x + 2(i-1)) \bullet 2iH(i-1; x) \\
	&= (\partial_x^2 -2x\partial_x + 2(i-1)) \bullet \partial_x H(i; x). \label{eq:ddodhp1}
\end{align}
% is obtained. 
On the other hand, 
$(\shift_i^2 - 2x\shift_i + 2i) \bullet H(i-1; x) = 0$ holds from~\eqref{eq:ddohp2}.
% \begin{equation}
% 	(\shift_i^2 - 2x\shift_i + 2i) \bullet H(i-1; x) = 0 
% \end{equation}
By multiplying the both sides by $2(i+1)(i+2)$ from the left, we obtain
\begin{align}
	0 &= 2(i+1)(i+2)(\shift_i^2 - 2x\shift_i + 2i) \bullet H(i-1; x) \\
	&= ((i+1)\shift_i^2 - 2x(i+2)\shift_i + 2(i+1)(i+2)) \\
	&\qquad \qquad \qquad \qquad \qquad \qquad\bullet 2iH(i-1; x), \\
	&= ((i+1)\shift_i^2 - 2x(i+2)\shift_i + 2(i+1)(i+2)) \\
	&\qquad \qquad \qquad \qquad \qquad \qquad\bullet \partial_xH(i; x), \label{eq:ddodhp2}
\end{align}
where the equalities $(i+2)\shift_i^2 = \shift_i^2i$ and $(i+1)\shift_i = \shift_i i$ are used for deriving the second line.
Eventually, we obtain the differential operators that annihilates $\mathscr{Z}[\partial_x H](s, x)$ as the inverse Mellin transform of~\eqref{eq:ddodhp1} and~\eqref{eq:ddodhp2}, that is, 
\begin{align}
	&\hat{\mu}(\partial_x^2 -2x\partial_x + 2(i-1)) = \partial_x^2 -2x\partial_x - 2s\partial_s - 4, \label{eq:dodhp1} \\
	&\hat{\mu}((i+1)\shift_i^2 - 2x(i+2)\shift_i + 2(i+1)(i+2)) \\
	% =& (-s\partial_s)s^2 - 2x(-s\partial_s + 1)s + 2(-s\partial_s)(-s\partial_s + 1) \\
	&= 2s^2\partial_s^2 + (2s^2x-s^3)\partial_s-2s^2. \label{eq:dodhp2}
\end{align}
% \begin{equation}
% 	\hat{\mu}(\partial_x^2 -2x\partial_x + 2(i-1)) = \partial_x^2 -2x\partial_x - 2s\partial_s - 4 \label{eq:dodhp1}
% \end{equation}
% and
% \begin{align}
% 	&\hat{\mu}((i+1)\shift_i^2 - 2x(i+2)\shift_i + 2(i+1)(i+2)) \\
% 	=& (-s\partial_s)s^2 - 2x(-s\partial_s + 1)s + 2(-s\partial_s)(-s\partial_s + 1) \\
% 	=& 2s^2\partial_s^2 + (2s^2x-s^3)\partial_s-2s^2. \label{eq:dodhp2}
% 	% &(\partial_x^2 -2x\partial_x - 2s\partial_s - 4) \bullet \widetilde{\partial_x H}(s, x) = 0, \\
% 	% &((-s\partial_s)s^2 - 2x(-s\partial_s + 1)s + 2(-s\partial_s)(-s\partial_s + 1)) \bullet \widetilde{\partial_x H}(s, x)
% \end{align}
We can confirm that the differential operators in~\eqref{eq:dodhp1} and~\eqref{eq:dodhp2} generate a holonomic ideal by using a CAS and prove the following lemma. 
\begin{lemma}
	The z-transform $\mathscr{Z}[\partial_x H](s, x)$ is holonomic. 
\end{lemma}

As each factor of the integrands in~\eqref{eq:ZT1}--\eqref{eq:ZT3} is holonomic, the integrands are also holonomic (Lemma~\ref{lem:holprod}). 
Here, the following assumption is made. 
\begin{assumption} \label{assm:rapid}
	All the integrands in~\eqref{eq:ZT1}--\eqref{eq:ZT3} are rapidly decreasing with respect to $x$. 
\end{assumption}

Note that Assumption~\ref{assm:rapid} holds as long as $f$, $g$, and $q$ are of exponential growth. 
Hence, \eqref{eq:ZT1}--\eqref{eq:ZT3} are holonomic (Lemma~\ref{lem:holinteg}), and the differential operators annihilating them can be computed by symbolic computation~\cite{Oaku2003}. 
% Moreover, if the integrand is rapidly decreasing, $\mathscr{Z}[a](s, t, v)$ is holonomic and differential operators that annihilates $\mathscr{Z}[a]$ can be computed~\cite{Oaku2003}. 
% According to the weighting function $\exp(-x^2)$, the integrands are rapidly decreasing as long as the functions $f$, $g$, and $q$ are of exponential growth. 
Consequently, recurrence relations satisfied by $a(i,j,k)$, $b(i,k)$, and $c(k)$ can be obtained from the differential operators via the Mellin transform~\eqref{eq:MT}. 
The whole procedure is summarized in Algorithm~\ref{alg:RecRel}. 
\begin{algorithm}[b]
	\caption{Derivation of recurrence relations satisfied by $a[i,j,k]$, $b[i,k]$, and $c[k]$\label{alg:RecRel}}
	\begin{algorithmic}[1]
		\Require{Sets of differential operators $\dop{G}_f, \dop{G}_g, \dop{G}_q \subset \D_1$ that annihilate $f(x), g(x), q(x)$, respectively, and $\dop{G}_H, \dop{G}_{\partial H} \subset \D_2$ that annihilate $\mathscr{Z}[H](s, x), \mathscr{Z}[\partial_x H](s, x)$, respectively}
		\Ensure{Set of difference operators $\sop{G}_a \subset \Ore_3$, $\sop{G}_b \subset \Ore_2$, and $\sop{G}_c \subset \Ore_1$ that annihilate $a(i,j,k)$, $b(i,k)$, and $c(k)$, respectively}
		% \State{Compute sets of differential operators $\tilde{I}_H, \tilde{I}_{\partial H} \subset \D_{2}$ from $I_H, I_{\partial H}$ via $\hat{\mu}$, respectively}
		% \If{Any one of ideals generated by $\tilde{I}_H$ and $\tilde{I}_{\partial H}$ in $\D_2$ is not holonomic}{\ terminate}
		% \EndIf
		\State{Compute sets of differential operators $\dop{J}_a \subset \D_4$, $\dop{J}_b \subset \D_3$, and $\dop{J}_c \subset \D_2$ that annihilate the integrands of~\eqref{eq:ZT1}, \eqref{eq:ZT2}, and~\eqref{eq:ZT3}, respectively, from inputs and $\partial_x^2 + 2x$ that annihilates $\exp(-x^2)$}
		\State{Compute sets of differential operators $\dop{K}_a \subset \D_3$, $\dop{K}_b \subset \D_2$, and $\dop{K}_c \subset \D$ that annihilates $\mathscr{Z}[a]$, $\mathscr{Z}[b]$, and $\mathscr{Z}[c]$ from $\dop{J}_a$, $\dop{J}_b$, and $\dop{J}_c$, respectively}
		\State{Compute sets of difference operators $\sop{G}_a$, $\sop{G}_b$, and $\sop{G}_c$ from $\dop{K}_a$, $\dop{K}_b$, and $\dop{K}_c$, respectively, via Mellin transform~\eqref{eq:MT}}
	\end{algorithmic}
\end{algorithm}

\section{Numerical Example}

This section presents a numerical example to show the efficiency of the proposed method. 
In the following demonstration, \texttt{Risa/Asir} and \texttt{Julia} are used to perform symbolic computation and numerical computation, respectively, of the proposed method. 
In addition, \texttt{Maple} is used to perform an existing numerical integration method for comparison. 
All computations are performed on a PC(Intel(R) Core(TM) i9-10920X CPU @ 3.50GHz; RAM: 64 GB). 
% All computations are performed on a PC(). 
% Algorithms~\ref{alg:RecRela} and~\ref{alg:RecRelb} are implemented

\subsection{Problem setting}
Consider the nonlinear optimal control problem of the nonlinear scalar system:
\begin{equation}
	\dot{x} = \sin(x) + u(x), \quad J = \int_0^{\infty} \frac{1}{2} x^2(t) + \frac{1}{2} u^2(t) dt. \label{eq:ex_problem}
\end{equation}
The nonlinear functions $\sin(x)$ and $x^2$ are holonomic, which are annihilated by differential operators $\partial_x^2 + 1$ and $\partial_x^3$, respectively. 
% with the performance index
% \begin{equation}
% 	V = \int_0^{\infty} \frac{1}{2} x^2(t) + \frac{1}{2} u^2(t) dt. 
% \end{equation}
% The HJB and GHJB equations for this problem are obtained as follows.
% \begin{align}
% 	&\mathrm{HJB}(V) = \pdev{V}{x}(x)\sin(x) + \frac{1}{2}x^2 - \frac{1}{2}\left\{ \pdev{V}{x}(x) \right\}^2 = 0, \\
% 	&\mathrm{GHJB}(V, u) = \pdev{V}{x}(x)\left(	\sin(x) + u \right) + \frac{1}{2}x^2 + \frac{1}{2}u^2 = 0. \label{eq:exGHJB}
% \end{align}

To perform the SGA (Algorithm~\ref{alg:SGA}), the following integrals need to be computed: 
% \begin{equation}
% 	\begin{gathered}
% 	a(i,j,k) = \left\langle 2\partial_x H(i; x) \partial_x H(j; x), H(k; x) \right\rangle, \\
% 	b(i,k) = \left\langle \partial_x H(i; x)\sin(x), H(k; x) \right\rangle, \quad c(k) = \frac{1}{2}\left\langle x^2, H(k; x) \right\rangle. 
% 	\end{gathered}
% \end{equation}
\begin{align}
	a(i,j,k) &= \left\langle 2\partial_x H(i; x) \partial_x H(j; x), H(k; x) \right\rangle, \\
	b(i,k) &= \left\langle \partial_x H(i; x)\sin(x), H(k; x) \right\rangle, \\
	c(k) &= \frac{1}{2}\left\langle x^2, H(k; x) \right\rangle. 
\end{align}

\subsection{Symbolic part}
To obtain the recurrence relations satisfied by $a(i,j,k)$, $b(i,k)$, and $c(k)$, Algorithm~\ref{alg:RecRel} can be applied to the inputs $\dop{G}_f = \{\partial_x^2 + 1\}$, $\dop{G}_g = \{0\}$, $\dop{G}_q = \{\partial_x^3\}$, $\dop{G}_H$ being the set of~\eqref{eq:dohp1} and~\eqref{eq:dohp2}, and $\dop{G}_{\partial H}$ being the set of~\eqref{eq:dodhp1} and~\eqref{eq:dodhp2}. 
For example, $\dop{J}_a \subset \bm{R}[x, s, t, v]\langle \partial_x, \partial_s, \partial_t, \partial_v \rangle$ is obtained as the set of 19 differential operators including 
\begin{align}
	% &2v\partial_v + 2vx - v^2, \\
	&\partial_x + 2s\partial_s^2 + (2sx-s^2+2)\partial_s \\
	&\quad + 2t\partial_t + (2tx - t^2 + 2)\partial_t+v-2s-2t, \label{eq:exJop}
\end{align}
where the remaining elements are omitted due to the space limit. 
It can be seen that \eqref{eq:exJop} is invariant under the interchange of $s$ and $t$, which is consistent with the integrand of~\eqref{eq:ZT1} being invariant under the interchange of $i = \mu(-s \partial_s -1)$ and $j = \mu(-t \partial_t -1)$. 
% As the output of Algorithm~\ref{alg:RecRela}, we obtain $\sop{G}_a$ as listed in Table~\ref{tab:ex_SGAsops}. 
% As the output of Algorithm~\ref{alg:RecRel}, we can obtain $\sop{G}_a$ as the following three difference operators: 
As the output of Algorithm~\ref{alg:RecRel}, we obtain $\sop{G}_a$, $\sop{G}_b$, and $\sop{G}_c$ as listed in Table~\ref{tab:ex_SGAsops}. 
\begin{table}[b]
	\centering
	\caption{Difference operators annihilating integrals $a(i,j,k)$, $b(i,k)$, and $c(k)$\label{tab:ex_SGAsops}}
	\renewcommand{\arraystretch}{1.3}
	\begin{tabular}{cp{7.5cm}} \hline\hline
		 \multirow{3}{*}{$\sop{G}_a$} &$(j +1) (-j +i +k +1)\shift_{i}+(i +1) (-j +i -k -1) \shift_{j}$ \\ %\cline{2-2}
		& $(k +1) (-k +i +j )\shift_{j}-(j +2) (-j +i +k )\shift_k$ \\ %\cline{2-2}
		&$(-j +i +k +1) (-k -3-j +i )\shift_{k}^{2} +4 (k +2) (k +1) (i +j -k -1)$ \\ \hline
		\multirow{3}{*}{$\sop{G}_b$}& $i (2 i -2 k -5)\shift_i \shift_k + (i +1) \shift_{k}^{2} - 4 (k +1) (i +1) (i -k -2)$\\ %\cline{2-2}
		 & $i (2 i -2 k -5)\shift_i^2 + (i +2) (2 i -2 k +1) \shift_{k}^{2} + 4 (i +2) (-k -1+i ) (2 i^{2}-4 i k +2 k^{2}-5 i +3 k -2)$\\ %\cline{2-2}
		& $(k +1) (-2 k +2 i -7) \shift_i + i(i +1) (-k -1+i )\shift_k^3 + 4 (i+1)(i^{3}+(-3 k -7) i^{2}+(3 k^{2}+13 k +\frac{53}{4}) i -k^{3}-6 k^{2}-11 k -6) \shift_{k}$\\ \hline
		$\sop{G}_c$ & $\shift_k^3$ \\ \hline
	\end{tabular}
	\renewcommand{\arraystretch}{1}
\end{table}
Note that $c(k)$ for this numerical example can be computed analytically as
\begin{equation}
	c(k) = \begin{cases}
		% \frac{\sqrt{\pi}}{4} & (k = 0) \\
		\sqrt{\pi} & (k = 2) \\
		0 & (\mbox{otherwise}), 
	\end{cases} \label{eq:c_analytic}
\end{equation}
which obviously satisfies $\shift_k^3 \bullet c(k) = c(k+3) = 0$ for $k \in \bm{N}$. 

\subsection{Numerical part}
From the recurrence relations obtained from the output of Algorithm~\ref{alg:RecRel}, the sets of integrals $A_N$, $B_N$, and $C_N$ can be computed recursively. 
Note that for a certain set of indices, the coefficients of the shift operators $\shift_i$, $\shift_j$, and $\shift_k$ in each difference operator may vanish, which indicates that the difference operator does not imply any recurrence relation on the set of indices. 

For example, the recurrence relation obtained from the second element of $\sop{G}_a$ in Table~\ref{tab:ex_SGAsops}: 
\begin{multline}
	(k +1)(-k +i +j )a(i,j+1,k) \\- (j +2) (-j +i +k )a(i,j,k+1) = 0 \label{eq:ex_arr2}
\end{multline}
can be used to compute one of $a(i,j+1,k)$ and $a(i,j,k+1)$ from the other. 
However, when the indices $i$, $j$, and $k$ satisfy $i+j-k=0$, the first term of~\eqref{eq:ex_arr2} vanishes regardless of the value $a(i,j+1,k)$, which just indicates
$(j+2)(-j+i+k)a(i,j,k+1) = 0$, 
% \begin{equation}
% 	(j+2)(-j+i+k)a[i,j,k+1] = 0, 
% \end{equation}
and does not give any information about the other values of $A_N$. 
For this reason, to compute all values of $A_N$, its several elements must be computed by using a numerical integration method. 
In this example, the following ten values are required: 
$a(1,1,1) = a(1,1,2) = a(2,1,2) = a(1,2,2)= a(4,1,1) = a(4,1,2) = a(1,4,1) = a(1,4,2) = 0, a(2,1,1) = a(1,2,1) = 56.7$. 
% \begin{align}
% 	&a(1,1,1) = a(1,1,2) = a(2,1,2) = a(1,2,2) \\
% 	&= a(4,1,1) = a(4,1,2) = a(1,4,1) = a(1,4,2) = 0, \\
% 	&a(2,1,1) = a(1,2,1) = 56.7. 
% \end{align}
Similarly, to compute all values of $B_N$, the following eight values are required: 
$b(1,2) = b(2,1) = b(3,2) = b(4,1) = 0, b(1,1) = 2.76, \quad b(1,3) = -2.76, b(3,1) =24.8, \quad b(3,3) =107$. 
% \begin{align}
% 	&b(1,2) = b(2,1) = b(3,2) = b(4,1) = 0, \\
% 	&b(1,1) = 2.76, \quad b(1,3) = -2.76, \\
% 	&b(3,1) =24.8, \quad b(3,3) =107.
% \end{align}

\subsection{Results and comparison}
% For $N = 10, 15$, the computational time of recurrence relations (\eqref{eq:ex_asop1}--\eqref{eq:ex_asop3}, and those listed in Table~\ref{tab:ex_SGAsops}) and the values $A_N$ and $B_N$ is compared to that of an existing numerical integration method implemented in \texttt{Maple}. 
For $N = 10, 15$, the computational time of the proposed method is compared to that of an existing numerical integration method implemented in \texttt{Maple} with four significant digits. 
The computational times are summarized in Table~\ref{tab:Cptime}. 
\begin{table}[b]
	\centering
	\caption{Comparison of computational times [s] with numerical integration\label{tab:Cptime}}
	\begin{tabular}{cccc} 
		&& Proposed & Maple\\ \hline\hline
		% \begin{tabular}{c}Recurrence \\ relations \end{tabular} & & $47.7$ s & N/A\\
		\multicolumn{2}{c}{Recurrence relations} & $47.7$ & N/A\\
		\multirow{3}{*}{$N = 10$}& $a(i,j,k)$ & $1.42 \times 10^{-3}$ & $98.2$\\
		& $b(i,k)$ & $3.26 \times 10^{-4}$ & $1.43 \times 10^2$\\
		& Total & $47.7$ & $2.44 \times 10^2$ \\
		\multirow{3}{*}{$N = 15$}& $a(i,j,k)$ & $1.51 \times 10^{-2}$ & $2.28 \times 10^2$\\
		& $b(i,k)$ & $6.26 \times 10^{-4}$ & $3.39 \times 10^2$\\
		& Total & $47.7$ & $5.67 \times 10^2$ \\ \hline
	\end{tabular}
\end{table}
As can be seen, the total times of the proposed method to compute $A_N$ and $B_N$ are much smaller than those of the existing method for both cases of $N = 10, 15$. 
% In particular, the numerical part of the proposed method does not take only few milliseconds because it just consists of recursive substitutions. 
In particular, the numerical part of the proposed method is finished within one second because it just consists of recursive substitutions. 

\begin{figure}[b]
	\centering
	\includegraphics[width=0.9\columnwidth]{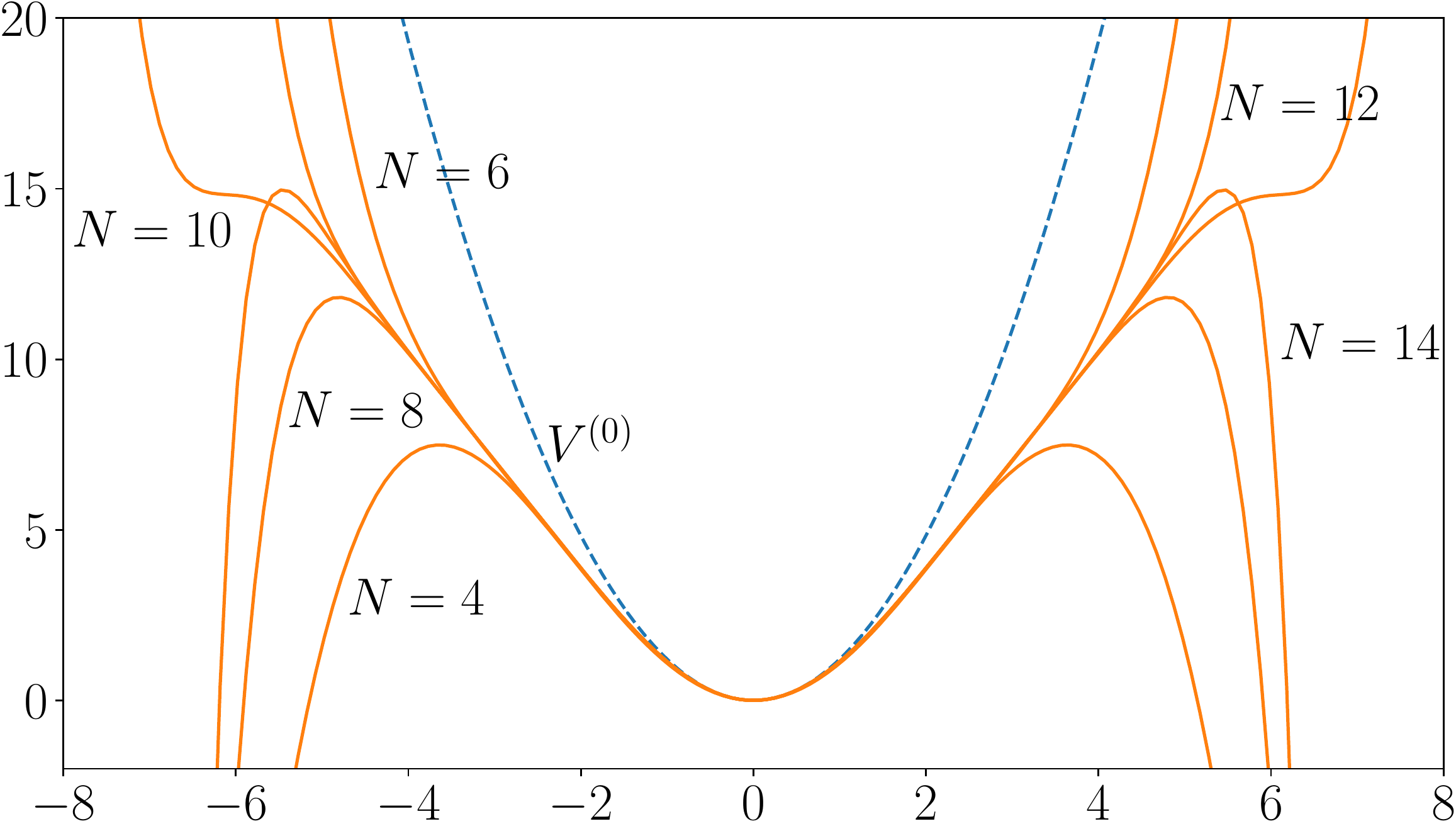}
	\caption{Value functions\label{fig:V}}
\end{figure}

\begin{figure}[b]
	\centering
	\includegraphics[width=0.8\columnwidth]{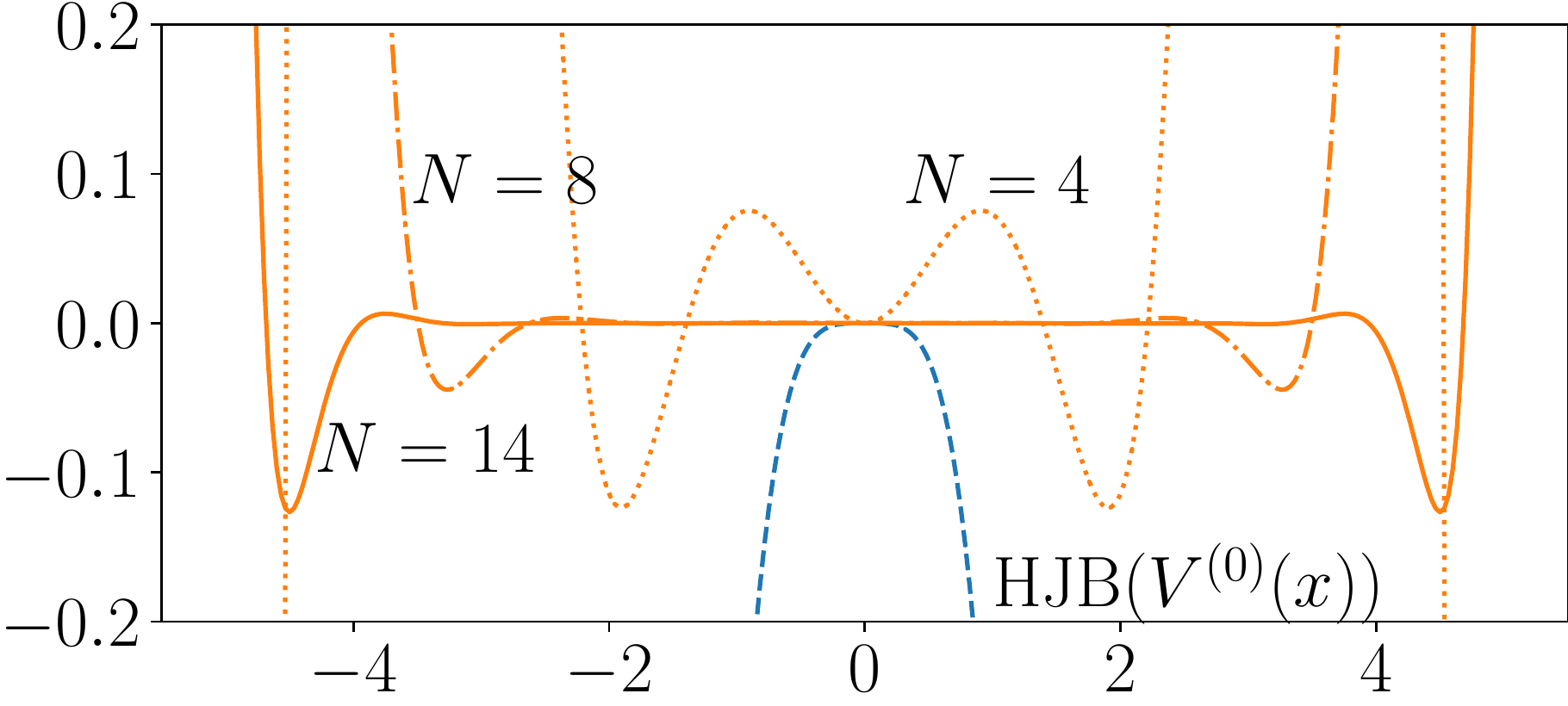}
	\caption{Evaluation of HJB equation\label{fig:HJB}}
\end{figure}

\begin{figure}[b]
	\centering
	\begin{minipage}{0.49\columnwidth}
		\centering
		\includegraphics[width=\columnwidth]{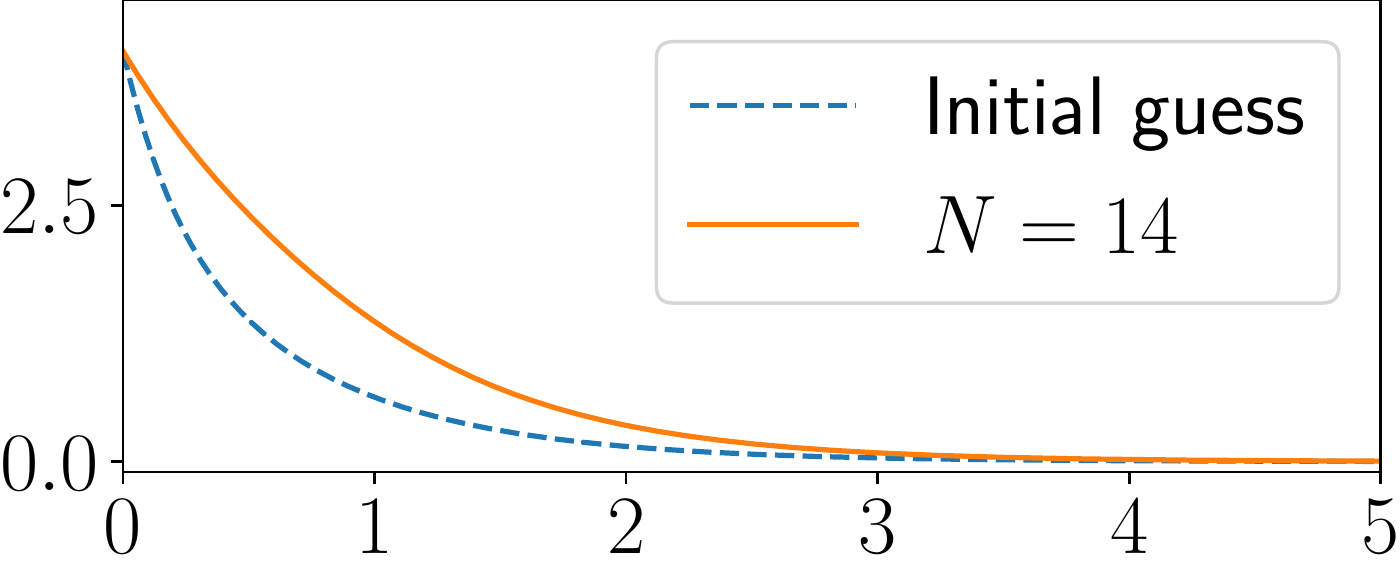}
		\subcaption{State}
	\end{minipage} \hfill
	\begin{minipage}{0.49\columnwidth}
		\centering
		\includegraphics[width=\columnwidth]{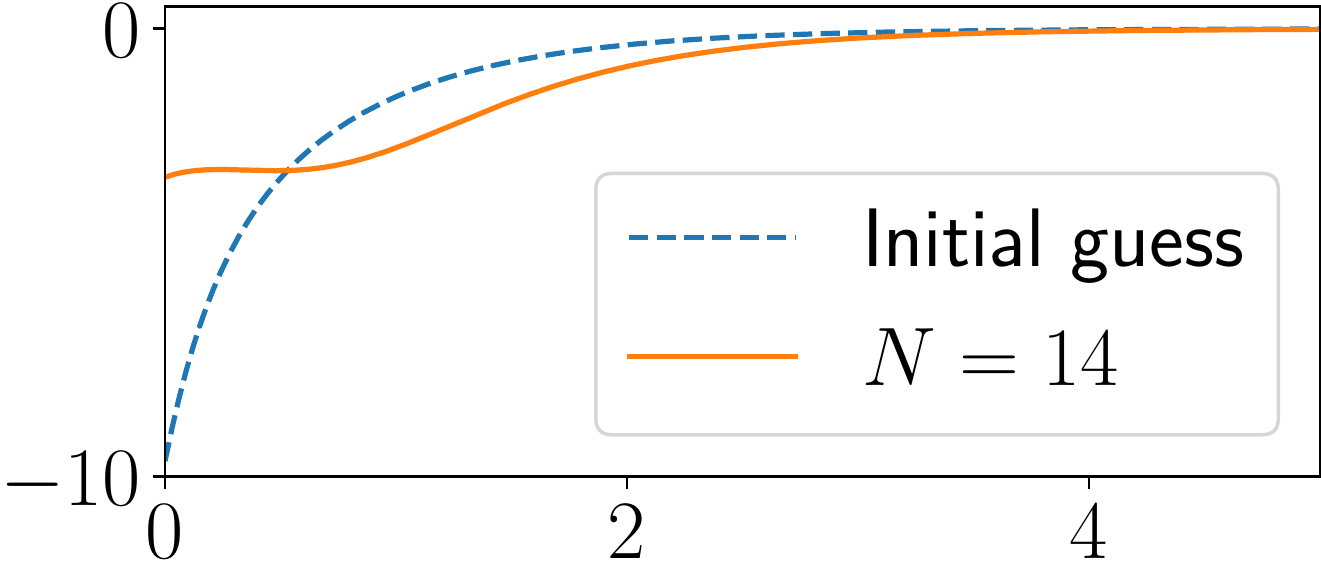}
		\subcaption{Input}
	\end{minipage}
	\caption{Trajectories with $x(0) = 4$\label{fig:sim}}
\end{figure}

The SGA (Algorithm~\ref{alg:SGA}) is performed with the integrals $A_N$, $B_N$, and $C_N$ for $N = 4, 6, \dots, 14$.
The initial guess $v^{(0)}$ is set to
\begin{equation}
	v^{(0)}_k = \begin{cases}
		\frac{1 + \sqrt{2}}{8} & (k=2) \\
		0 & (\mbox{otherwise})
	\end{cases}, 
\end{equation}
which corresponds to $V^{(0)}(x) = v^{(0)}_2H(2; x)$ and is equal to the value function $V(x) = (1+\sqrt{2})x^2/2$ of the linearized system $\dot{x} = x + u$ with $J$ in~\eqref{eq:ex_problem} up to a constant. 
Figure~\ref{fig:V} shows $V^{(0)}(x)$ and $V^*(x)$ obtained for $N = 4, 6, \dots, 14$, where the constant term of each value function is adjusted so that $V^*(0) = 0$.
In Fig.~\ref{fig:HJB}, $\mathrm{HJB}(V^{(0)}(x))$ and $\mathrm{HJB}(V^*(x))$ of the value functions for $N=4, 8, 14$ are depicted. 
As can be seen, the region where $\mathrm{HJB}(V^*(x)) \simeq 0$ glows as $N$ increases, which indicates that $A_N$ and $B_N$ are computed validly by the proposed method. 
Finally, the state and input trajectories with $x(0) = 4$ are illustrated in Fig.~\ref{fig:sim}, which shows that the feedback law obtained from $V^*(x)$ with $N=14$ stabilizes the system while suppressing the magnitude of the input. 

% Although these differential operators may give sufficient number of recurrence relations to compute the SGA parameters up to any order, it is not clear at a glance. 
% Hence, the differential operators in Table~\ref{tab:ex_SGAsops} must be transformed to more affordable forms called Pfaffian systems. 

% For example, the third element of $\sop{G}_a$ is equivalent to the following recurrence relation of a vector-valued function $q_a = [a\ \shift_ka]^\top$:  
% \begin{equation}
% 	\shift_k q_a = \begin{bmatrix}
% 		\shift_k a \\
% 		\shift_k^2 a
% 	\end{bmatrix} = \begin{bmatrix}
% 		0 & 1 \\
% 		-\frac{4 (k +2) (k +1) (i +j -k -1)}{(-j +i +k +1) (-k -3-j +i )} & 0
% 	\end{bmatrix} q_a
% \end{equation}

% All computations are performed on a PC(Intel(R) Core(TM) i9-10920X CPU @ 3.50GHz; RAM: 64 GB). 

% By using these recurrence relations, the SGA parameters $a[i,j,k]$, $b[i,k]$, and $c[k]$ can be computed recursively.

% As the input of Algorithm~\ref{alg:RecRela}, 
% The holonomic functions and sequences are defined through the differential and difference operators that annihilates them. 
% More specifically, the ideal of differential operators that annihilates a holonomic function is defined as follows. 
% To this end, we need to introduce the notion of \textit{holonomic functions and sequences}. 
% Since the 

\section{Conclusion}
% Note that the arguments in Section~\ref{sec:RecRel} can be readily extended to multi-dimensional cases.  
In this paper, a symbolic-numeric computation method for the integrals in the successive approximation algorithm has been proposed.
The successive approximation algorithm approximates a solution of Hamilton-Jacobi-Bellman equation by iteratively solving a linear equation, which is defined with multiple integrals of nonlinear functions. 
By using the symbolic computation of differential operators and the Mellin transform of difference-differential operators, a set of recurrence relations satisfied by the integrals can be computed. 
From a finite number of the integrals computed by numerical integration, the other integrals can be computed up to any number by using the recurrence relations. 

% For future work, the proposed method will be extended to apply it to multi-dimensional nonlinear systems. 
Future work directions of this study includes the extension of the proposed method to the case of multi-dimensional systems. 
Moreover, the approximation with another orthogonal polynomial such as the Chebyshev polynomials could be included. 

% \bibliographystyle{jecon}
% \bibliography{library}
%%%%%%%%%%%%%%%%%%%%%%%%%%%%%%%%%%%%%%%%%%%%%%%%%%%%%%%%%%%%%%%%%%%
%
%  This bbl file is created by jecon.bst ver.5.5
%  The latest jecon.bst is available at
%  <http://shirotakeda.org/ja/tex-ja/jecon-ja.html>
%
%%%%%%%%%%%%%%%%%%%%%%%%%%%%%%%%%%%%%%%%%%%%%%%%%%%%%%%%%%%%%%%%%%%

\end{document}